# Exploring the impact of mobility restrictions on the COVID-19 spreading through an agent-based approach

Martina Fazio[1*], Alessandro Pluchino[1,2], Giuseppe Inturri[3], Michela Le Pira[4], Nadia Giuffrida[5], Matteo Ignaccolo[4]

[1]Department of Physics and Astronomy, University of Catania, Catania, Italy
[2]INFN Section of Catania, Catania, Italy
[3]Department of Electrical, Electronic and Computer Engineering, University of Catania, Catania, Italy
[4]Department of Civil Engineering and Architecture, University of Catania, Catania, Italy
[5]Spatial Dynamics Lab, University College Dublin, UCD Richview Campus, D04 V1W8, Belfield, Dublin, Ireland

*Corresponding author: Martina Fazio, e-mail: martina.fazio@phd.unict.it

**Abstract**
Mobility restriction is considered one of the main policies to contain COVID-10 spreading. However, there are multiple ways to reduce mobility via differentiated restrictions, and it is not easy to predict the actual impact on virus spreading. This is a limitation for policy-makers who need to implement effective and timely measures. Notwithstanding the big role of data analysis to understand this phenomenon, it is also important to have more general models capable of predicting the impact of different scenarios. Besides, they should be able to simulate scenarios in a disaggregated way, so to understand the possible impact of targeted strategies, e.g. on a geographical scale or in relation to other variables associated with the potential risk of infection. This paper presents an agent-based model (ABM) able to dynamically simulate the COVID-19 spreading under different mobility restriction scenarios. The model uses the Italian case study with its 20 administrative regions and considers parameters that can be attributed to the diffusion and lethality of the virus (based on a virus spread risk model) and population mobility patterns. The model is calibrated with real data and reproduces the impact that different mobility restrictions can have on the pandemic diffusion based on a combination of static and dynamic parameters. Results suggest that virus spreading would have been similar if differentiated mobility restriction strategies based on *a-priori* risk parameters instead of a national lockdown would have been put in place in Italy during the first wave of the pandemic. The proposed model could give useful suggestions for decision-makers to tackle pandemics and virus spreading at a strategic level.

## 1. Introduction

The recent health emergency caused by the COVID-19 pandemic has forced people to change their mobility behaviours, with the reduction of leisure travels and the promotion of teleworking and online educational activities (de Vos, 2020). Among the most applied contagion control measures, those relating to the limitation of travels, the so-called "stay-at-home" order, have become widespread with the aim of avoiding the circulation of the virus in public environments. Public transport has been highly impacted both by government restrictions and travellers' choices (Jenelius and Cebecauer, 2020; Gutiérrez et al. 2020).

In the acute phases of the emergency, the disease has forced government agencies to consider several preventive measures to control its spreading. In Italy, during the first wave of 2020 a national lockdown of about two months was imposed by the government to limit population mobility, provoking a reduction in urban travel and number of air flights. During the second wave started in fall 2020, differentiated strategies have been implemented according to the "colours" of the regions based on multiple sanitary indicators (Ministero della Salute, 2020). However, data are not easily accessible or clearly explained to the public, resulting in uncertainties and, sometimes, leading to protests from the regional government departments. Notwithstanding the effectiveness of social distancing measures, the debate on the actual impact of travel limitation measures is very lively, both in the academic world and in public opinion. Public transport is a case in point, being its influence on the virus spreading debated (Tirachini and Cats, 2020).

Mobility restrictions indeed affect the economic conditions of both people and governments and are also responsible for a "segregation effect" of people with low income (Porcelli et al., 2020). It is therefore important to quantify the effectiveness of various measures on the spread of the virus, to avoid overestimating the effects of health prevention that can generate various equally serious economic and social externalities.

Recent literature tries to quantify the actual effects of these restrictions on the virus spreading, e.g. via a multiple linear regression model also including the number of tests/day and environmental variables (Cartenì et al., 2020). However, the use of data on the spread of the infection in terms of total number of cases can lead to incorrect analysis results, precisely because this value is conditioned by the actual number of swabs carried out on the population. Indeed, the quantification of the results of these restrictions is difficult, especially due to the biased data available.

Some authors focused on data of excess deaths, which are likely to be less affected by specific assumptions, and correlated them with mobile data, showing that mobility is responsible for more than 90% of the initial spreading in Italy and in France (Iacus et al., 2020). Other studies referred to the USA and Chinese cases analysed huge amounts of data and statistical models were used to show the strong correlation of such restrictions on the virus spreading (Xiong et al., 2020; Badr et al., 2020; Kraemer et al., 2020).

Data analysis techniques usually play a big role to understand this type of phenoma; however, due to the rapid development of the pandemic, it is also important to have more general models capable of predicting the impact of different scenarios, independent from the available infection data. Simulation models could help to understand the possible impact of differentiated strategies (e.g. according to the geographical scale), and replicate the related scenarios in a disaggregated way.

In this respect, agent-based models (ABM) have many advantages, among them the possibility of having a very rich data scenario with country-specific demography, the possibility of simulating complex social interactions and population mobility patterns. A further advantage of using the ABM approach is the stochastic nature of the simulations, which allows to implement a component of randomness (Huppert & Katriel, 2013; Shi et al., 2014). ABM have been used to simulate virus spreading, drawing inspiration from the so called SIR-based models (i.e. susceptible–infective-removed) and applying them to a dynamic simulation environment. Silva et al. (2020) proposed the so called COVID-ABS to simulate different scenarios (e.g. lockdown, use of face masks) and estimated the economic impact of them. However, they do not apply the model in a real case study, but reproduce a synthetic population of a closed society. Najmi et al. (2020) extended an existing activity-based model named SydneyGMA model to replicate the case of Sydney by determining COVID-19-specific parameters and considering the interaction among agents and, thus, resulting in a useful model at a city level.

This paper proposes an agent-based model to dynamically simulate the impact of mobility restrictions on the spreading of the COVID-19 at a national scale. The model proposed is new since it reproduces a real case study, i.e. Italy, at the level of details of country regions and considers multiple data sources and *a-priori* parameters that can be related to the risk of spreading. To build the ABM, we drew inspiration from a previous study aimed at measuring an *a-priori* risk index for each of the 20 regions in Italy (Pluchino et al., 2021). The Authors showed that the geographic distribution of this index correlates with the available COVID-19 official data related to the pandemic spreading. Based on this, possible policy interventions have been suggested to tackle the virus spreading. In this study, a dynamical version of this model has been implemented in a simulation environment to test the previous findings through an agent-based approach. In addition, a scenario analysis will be presented, differentiating region by region the measures to restrict mobility that could have been implemented to struggle the pandemic.

The remainder of the paper is organised as follows. Section 2 presents the data and methods used to build the ABM, while section 3 introduces the case study and the related model steps. Section 4 presents and discusses the results with some policy implications. Section 5 concludes the paper.

## 2. Data and Methods

The rationale behind the use of ABM is to evaluate how the epidemic spread changes on the basis of different mobility restriction policies.

To build the ABM, authors drew inspiration from a previous study aimed at measuring an *a-priori* risk index for each of the 20 regions in Italy (Pluchino et al., 2021). The study showed that the geographic distribution of this index correlates with the available COVID-19 official data about the number of infected individuals, patients in intensive care and the number of deaths. More in detail, the risk index was built combining the following indicators, extracted from data collected on a regional basis before the beginning of the pandemic: mean winter temperature (Wt), since low temperatures affect the spread and transmission of the virus; housing concentration (Hc), since urbanization of cities leads to a more threatening diseases diffusion; healthcare density (Hcd), as it was found the potential of hospitals to favour super-spreading events; population mobility (Pm), since this favour the interaction among people and the virus transmission; air pollution (Ap): the correlation between exposure to particulate pollution and the diffusion of COVID-19 is demonstrated by various studies; population over 60 (P_over60), considered more vulnerable to suffer virus effects. References and more detailed information can be found in Pluchino et al. (2021).

In the ABM proposed there are two type of agents: regions and individuals. Based on the actual population of each region, a proportional number of individuals is assigned, considering that each individual-agent is representative of a certain number of real individuals (with a scale of approximately 1:1000 based on the actual number of Italian population).

This approximation was made in order to avoid excessive simulation time arising from considering the real scale of the Italian population.

A description of the parameters affecting agents' behaviours is provided in section 3.1.

The construction of the model can be summarized in the following steps:
- agents setup;
- virus spreading model;
- scenario setup;
- scenario simulation.

The ABM simulations were carried out through the NetLogo software, which is a multi-agent programmable environment for simulating and modelling complex systems by taking into account the evolution of the "agents" over time (Wilensky, 1999).

In the following, the case study of Italy is presented more in detail together with a description of the related model's phases.

## 3. Case study

The case study analysed in this work is related to Italy and its 20 administrative regions. Italy was the first European country in which the virus appeared, although the dynamics of spread and the date of the first infection remain uncertain. The first confirmed cases of contagion date back to 2020, January 23rd, when two tourists from China were tested positive for the virus in Rome. The first two outbreaks of COVID-19 infections with positive cases of Italian citizens were reported later on February 21st, in Lombardy and Veneto[1]. Since then, the infection has spread throughout Italy with varying intensity. Nevertheless, several studies have shown that there were actually cases even before (Apolone et al., 2020, Valenti et al., 2020). On March 7th a government measure imposed some travel limitations. On March 11th, the restrictive measures were converted into a national lockdown, with a "stay-at-home" order allowing travelling only for essential services or urgent reasons, with the aim of stopping the spread of the virus. This national lockdown of about two months has provoked a tough reduction both in short and long distance travelling.

Figure 1a shows statistics on the number of daily cases in Italy. As can be seen, the trend increases starting from March and it seems to have a surge in this second wave starting from October 2020. The first trend is justified by the difficulty to accurately detect the actual number of infected (Tradigo et al., 2020). Subsequently, with the growth in the number of swabs, the share of recorded infected people has gradually increased. Nevertheless, uncertainties on the actual number of cases still remain, due to the biased data available on the contagion rate.

In October 2020, the World Health Organization (WHO) stated that 10 percent of the global world population was infected with the virus[2]. This leads to the belief that also in Italy the number of infected was actually much higher than reported by official data sources, touching the millions of infected. This number is also comparable with the average annual number of seasonal flu cases[3]. The number of daily deaths (Figure 1b) has instead the same order of magnitude in both waves.

The absence of reliable data on the number of infections did not allow to have clear information on the actual effects of the restrictions imposed in the first COVID-19 wave on the virus spreading.

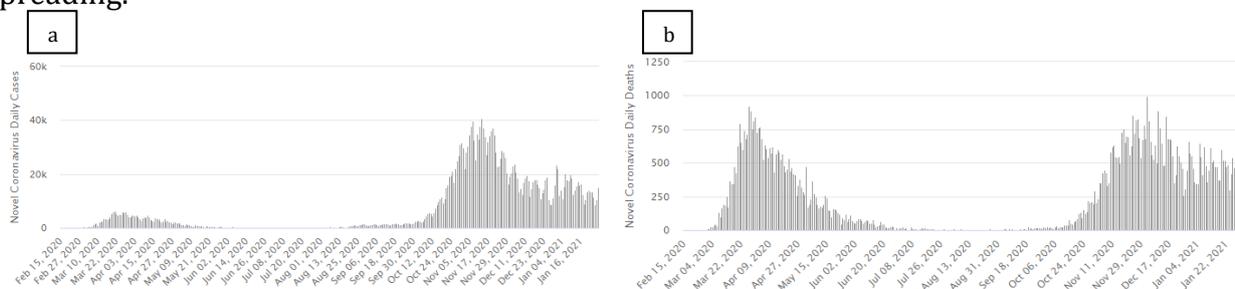

*Figure 1 - Daily New Cases (a); Daily Deaths (b) in Italy (source: https://www.worldometers.info/coronavirus/country/italy/)*

---

[1] https://lab24.ilsole24ore.com/storia-coronavirus/

[2] https://www.cnbc.com/2020/10/05/who-10percent-of-worlds-people-may-have-been-infected-with-virus-.html

[3] https://www.epicentro.iss.it/influenza/stagione-2019-2020-primo-bilancio

Based on these premises, we developed an ABM able to reproduce a contagion rate which matches WHO statistics (resulting in a higher number of total cases) and the differentiation of infections between the Italian regions visible from the data collected by the Italian Ministry of Health. In this paper the authors will use the model to simulate different mobility restriction scenarios and evaluate the related impacts during the first epidemic wave in 2020. As shown in Figure 2, the realistic geographical distribution of representative agents on the Italian territory allows to simulate the population mobility either in absence or in presence of restrictions, evaluating in real-time the virus diffusion and the corresponding effects in terms of infections and mortality.

Thanks to the results obtained, it will be possible to provide suggestions on mobility restrictions for an emergency plan that could be adapted not only for the case of COVID-19, but also for other pandemics.

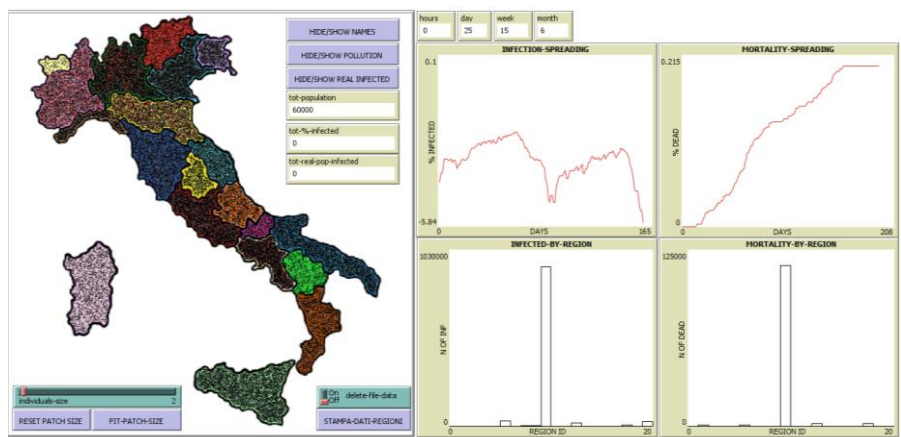

*Figure 2 - Simulation environment*

The steps of the ABM and the selected scenarios are described in the following.

### 3.1. Model Steps

**Agents setup**

In the setup phase all the parameters relating to both regions and individuals are set. Each agent r (region) is characterized by the following parameters: Wt, Hc, Hcd, Ap, Pm and P_over60. Each agent i (representing individuals) inherits the first 5 parameters from its home region and is classified in an age-group depending on the percentage of P_over60.

The dynamics of the model is given by the changes in the Pm parameter (Pm_reduction), which are evaluated for each agent r and different time windows, according to mobility restriction.

Individual-agents in the model move according to the assigned mobility index multiplied by a variable probability of making a trip belonging to distances classes from 2 km to 50 km or more. (Figure 3). Trips over 50 km are considered as done by plane . For the mobility procedure reference was made from an Italian mobility report which provide an overview of the mobility habits of people in Italy ("16° Rapporto sulla mobilità degli italiani", source: ISFORT[4]). For airline mobility, authors referred to a dataset containing Origin-Destination matrix of airline travel for each region (source: ENAC, 2019[5]).

---

[4] https://www.isfort.it/progetti/16-rapporto-sulla-mobilita-degli-italiani-audimob/

[5] https://www.google.com/search?q=enac+dati+traffico+2019/

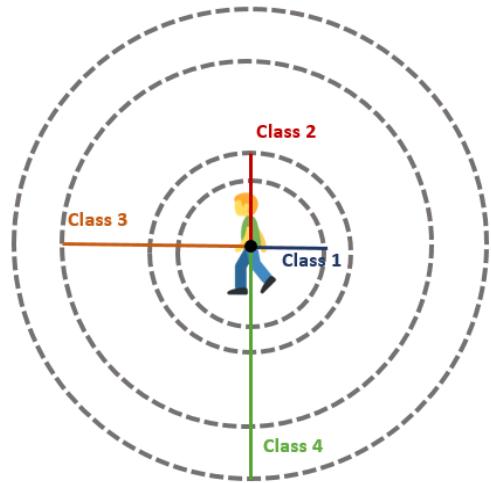

*Figure 3 - Scheme of distances classes*

The model is capable of dynamically reproduce mobility restrictions by using three main datasets that provided from March 7th until June 25th the reduction of air flights, daily mobility radius, and number of trips. OpenData are used to gather information about mobility decrease considering, in particular, the reduction of air flights from 2019 to 2020 and the reduction of the radius and number of overall trips (references are reported in Table 1).

Table 1 summarizes agents parameters (P). Each parameter has been normalized between 0 and 1, as in Pluchino et al. (2021).

| P | Description and unit | Source | Type |
|---|---|---|---|
| Wt | Average winter temperature (°C) | Italian Ministry of Agriculture (2016-2017) | Fixed for each region |
| Hc | Ratio between the total number of houses and the number of houses classified as "detached houses" | Italian Ministry of Economic Policy Planning and Coordination (2011) | Fixed for each region |
| Hcd | Number of hospital beds per inhabitant | Italian Ministry of Health (2019) | Fixed for each region |
| Ap | Exposure to concentrations of particulate matter (PM) | WHO (2016) | Fixed for each region |
| Pm | Ratio between the sum of commuting flows (incoming and outgoing) for a region and the population employed in the region. | Italian Ministry of Economic Policy Planning and Coordination (2011) | Fixed for each region |
| Pm_reduction | 1. reduction of air flights (%); 2. reduction of the dimension of the daily mobility radius (%); 3. reduction of the number of trips (%); | 1. EUROCONTROL; 2. Covid19mm.github: first report; 3. Google: covid19-mobility | Dynamic time windows |
| P_over60 | Fraction of population over 60 | ISTAT (2011) | Fixed for each region |

*Table 1 - Summary of agents parameters*

**Virus spreading model**

For the calculation of risk index (RI) authors referred to the Crichton's Risk Triangle (Crichton, D., 1999), which evaluates RI as a function of three parameters: hazard, vulnerability, and exposure. (i) Hazard takes into consideration those factors that can intervene in the spread of the infection; (ii) Vulnerability is a measure of an individual's likelihood of being infected; (iii) Exposure refers to the number of exposed people.

In the study of Pluchino et al. (2021), the RI is calculated for each region r as a floating point variable between 0 and 1 and is obtained as:

$$RI = HAZARD \cdot VULNERABILITY \cdot EXPOSURE \qquad (1)$$

Hazard, vulnerability and exposure are also floating point variables between 0 and 1, in turn calculated as follows:

$$HAZARD = 1/3 \cdot Hc + 1/3 \cdot Hcd + 1/3 \cdot Pm \qquad (2)$$

$$VULNERABILITY = 1/3 \cdot Wt + 1/3 \cdot Ap + 1/3 \cdot P\_over60 \qquad (3)$$

$$EXPOSURE = \text{population of each region} \qquad (4)$$

In the ABM model authors propose a dynamic version of RI by referring it to each individual-agents. In this respect the new risk index (ri) is calculated as follow:

$$ri = hazard \cdot vulnerability \qquad (5)$$

$$hazard = 1/3 \cdot Hc + 1/3 \cdot Hcd \qquad (6)$$

$$vulnerability = 1/3 \cdot Wt + 1/3 \cdot Ap \qquad (7)$$

The model provides a disaggregate version of RI in which the Pm, P-over60 and exposure component are specific characteristics referred to each agent i and therefore are not considered for the direct calculation of the risk index.

In order to simulate the total Italian population (about 59433744 individuals at the beginning of 2020) we adopt 60000 agents i, each one representing 991 real individuals, then we distribute them at random inside the territory of each region (see the black dots in Figure 2), proportionally to the respective inhabitants.

RI is therefore assigned to each region and also characterizes each individual living in that region, as explained below. Official data show that 95% of people died in Italy due to COVID-19 were aged over-60[6]. For this reason, for the calculation of RI, a distinction was made between under 60 and over 60, increasing the probability of being exposed for the latter category.

By combining the RI with the mean infection duration, the model determines the status associated to each individual on the basis of a SIR-based approach (Kermack and McKendrick, 1927): susceptible, infected, isolated (or not isolated), immune and dead.

While hazard and vulnerability are parameters directly linked to each individual, virulence and lethality are related to the characteristic of the virus. Virulence, which corresponds to the contagiousness of the virus, is a fixed parameter. Its value has been chosen through a calibration procedure by reproducing different scenarios by varying virulence until obtaining results comparable to the real data in terms of number of deaths. Also for the lethality, which correspond to the mortality level of the virus, reference was made to real data.

---

[6] https://www.epicentro.iss.it/coronavirus/sars-cov-2-decessi-italia

Due to the data uncertainty, lethality value varies according to a Gaussian probability distribution with mean 0.02 and standard deviation 0.01, i.e. lethality oscillates around 2% (Russell et al., 2020).

The mean infection duration, based on official data (Italian Ministry of Health[7]), is considered equal to 10 days.

Individual-agents change their status according to the procedure summarized in the following flowchart below (Figure 4) and described in the following.

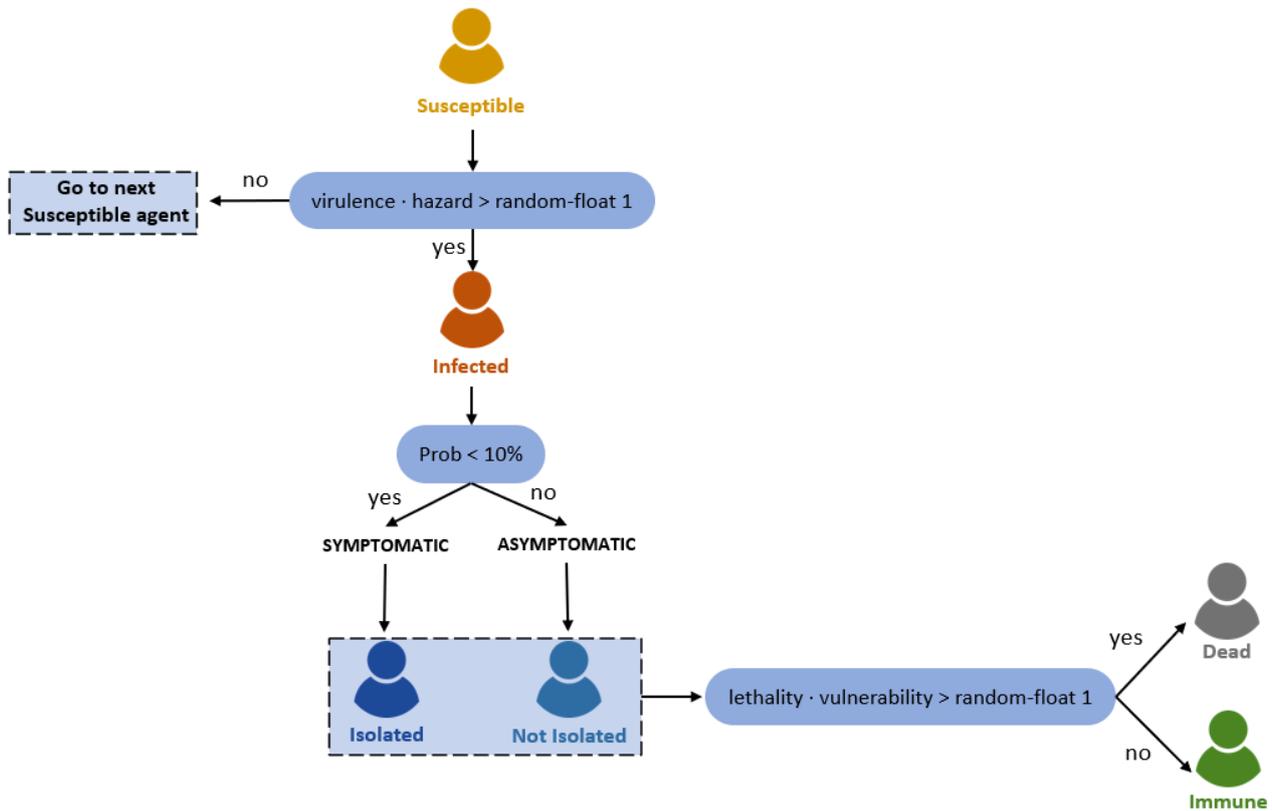

*Figure 4 - Individual's status change procedure flowchart*

The simulation starts with two agents with the status "infected" which represent the "zero patients". All the other individual-agents start from a status called "susceptible". Once the simulation starts and the individuals begin to travel according to the assigned mobility index, if a "susceptible" individual encounters an "infected" one, it will have a probability to contract the virus based on the product between hazard and virulence. If the result of this product is more than a random floating number between 0 and 1 (random-float 1 in Figure 4) , individuals will change their status to "infected". After getting infected, the individual is assigned with a probability of being symptomatic (≤ 10%) or asymptomatic (≥ 90%)[8], linked respectively to the new status or "isolated" and "not isolated". Finally, after the mean infection duration, the individual dies or recovers from the infection, by comparing the product between vulnerability and lethality with a random floating number between 0 and 1, assuming respectively "dead" or "immune" status.

**Choice of analysis scenarios**

---

[7] http://www.salute.gov.it/portale/nuovocoronavirus/dettaglioNotizieNuovoCoronavirus.jsp?lingua=italiano&id=5117
[8] https://www.istat.it/it/archivio/246156

According to policies suggestions by Pluchino et al. (2021), the following scenarios have been tested:
- *Status Quo*: Same total mobility restrictions for all regions;
- *Scenario N1*: No mobility restrictions at all (RI=1 for all regions);
- *Scenario S1*: Total mobility restrictions for all the regions (RI=1 for all regions);
- Different mobility restrictions according to 3 zones based on the values of the following parameters (Table 2, Table 3 and Figure 5):
    o mobility index (*Scenario M*);
    o hazard (*Scenario H*);
    o vulnerability (*Scenario V*);
    o risk index (*Scenario RI*).

More details on how these parameters were calculated can be found in Pluchino et al. (2021).

| Regions | mobility index | Zone | hazard | Zone | vulnerability | zone | risk index | Zone |
|---|---|---|---|---|---|---|---|---|
| Abruzzo | 0,15 | 1 | 0,33 | 1 | 0,06 | 1 | 0,04 | 1 |
| Basilicata | 0,03 | 1 | 0,23 | 1 | 0,01 | 1 | 0,01 | 1 |
| Calabria | 0,37 | 2 | 0,37 | 2 | 0,05 | 1 | 0,03 | 1 |
| Campania | 0,24 | 1 | 0,26 | 1 | 0,25 | 2 | 0,11 | 2 |
| Emilia Romagna | 0,81 | 3 | 0,69 | 3 | 0,23 | 2 | 0,27 | 2 |
| Friuli Venezia Giulia | 0,81 | 3 | 0,77 | 3 | 0,07 | 1 | 0,09 | 1 |
| Lazio | 0,53 | 2 | 0,50 | 2 | 0,23 | 2 | 0,19 | 2 |
| Liguria | 0,48 | 2 | 0,58 | 2 | 0,08 | 1 | 0,08 | 1 |
| Lombardia | 1,00 | 3 | 0,94 | 3 | 0,62 | 3 | 1,00 | 3 |
| Marche | 0,60 | 2 | 0,33 | 1 | 0,07 | 1 | 0,04 | 1 |
| Molise | 0,00 | 1 | 0,44 | 2 | 0,01 | 1 | 0,01 | 1 |
| Piemonte | 0,59 | 2 | 0,58 | 2 | 0,29 | 2 | 0,29 | 2 |
| Puglia | 0,29 | 1 | 0,45 | 2 | 0,11 | 2 | 0,09 | 1 |
| Sardegna | 0,29 | 1 | 0,60 | 2 | 0,04 | 1 | 0,04 | 1 |
| Sicilia | 0,54 | 2 | 0,53 | 2 | 0,09 | 1 | 0,08 | 1 |
| Toscana | 0,74 | 3 | 0,46 | 2 | 0,17 | 2 | 0,14 | 2 |
| Trentino Alto Adige | 0,66 | 2 | 0,74 | 3 | 0,04 | 1 | 0,05 | 1 |
| Umbria | 0,55 | 2 | 0,42 | 2 | 0,04 | 1 | 0,03 | 1 |
| Valle d'Aosta | 0,64 | 2 | 0,73 | 3 | 0,01 | 1 | 0,01 | 1 |
| Veneto | 0,95 | 3 | 0,69 | 3 | 0,27 | 2 | 0,32 | 2 |

Table 2 - Zone classification according to mobility index, hazard, vulnerability and risk index

| Risk index zone | Parameter | ZONE 1 | ZONE 2 | ZONE 3 |
|---|---|---|---|---|
| Scenario M | mobility index | No mobility restriction | 50% of mobility restriction | Total mobility restriction |
| Scenario H | hazard | | | |
| Scenario V | vulnerability | | | |
| Scenario RI | risk index | | | |

Table 3 - Characterization for zone-based scenarios

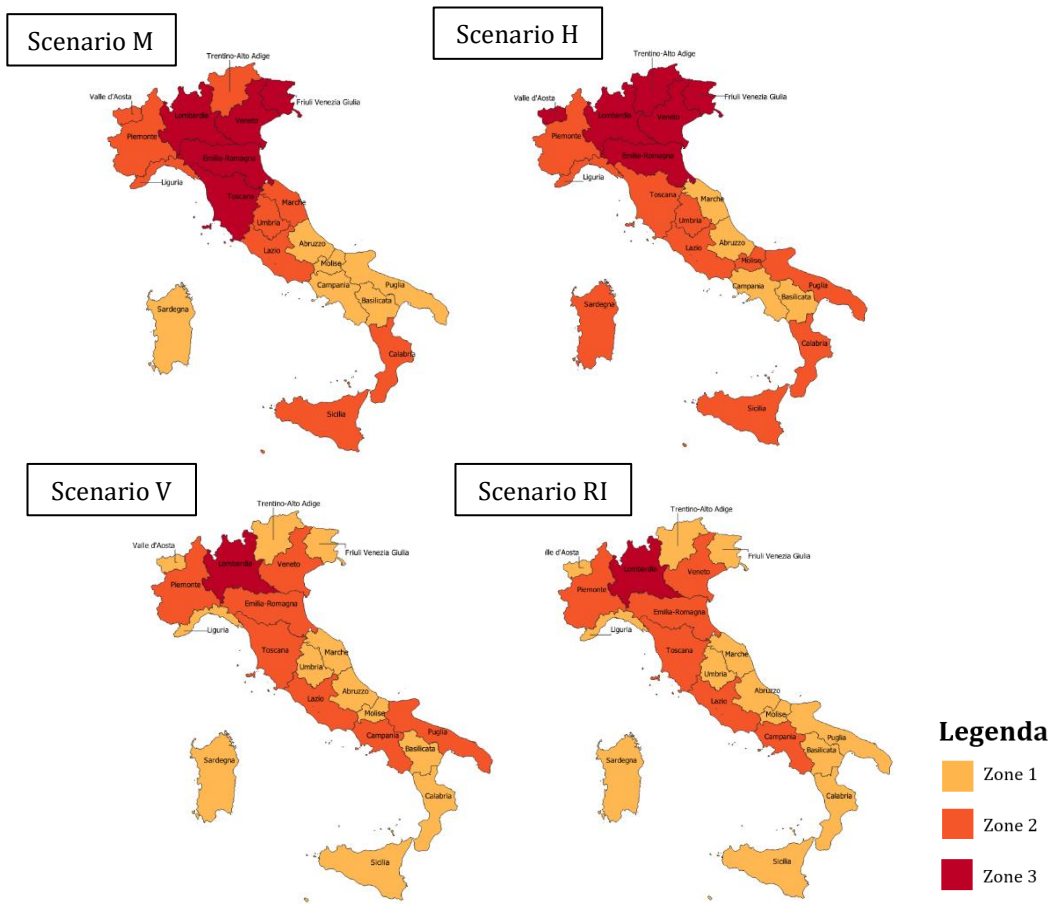

*Figure 5 – Classification in three zones, with increasing mobility restrictions, for scenarios defined in Table 3*

**Scenario simulation**

Following the real case study, the analysis of different scenarios has the same starting date as the governmental restriction introduced in Italy (i.e. March 7th, 2020). December 28th, 2019 was chosen as the starting pandemic date for all scenario, due to the uncertainty of the beginning of the infection in the country (Apolone et al., 2020, Valenti et al., 2020).

The two "zero patients" are in Lombardy and Lazio, regions where the first cases of COVID-19 occurred and also those where hub airports are present; hence they were considered the regions with more connections to other countries.

The output data of the simulation, calculated on June 25th, 2020 (after the end of the first epidemic wave), are the following:
 – Number of infected people for each region;
 – Number of dead people for each region.

Each scenario was simulated 5 times and the results were averaged to have a statistic of the events. Computing simulation time for each scenario assume reasonable values (about 20 minutes).

## 4. Results and discussion

In the following, results showed in Figure 6 and Figure 7 will be presented and discussed.
In the first rows of these figures, a comparison will be made between the *status quo* (a scenario assuming the national lockdown adopted by the Italian Government during the first epidemic wave) and the real COVID-19 data concerning, respectively, the cumulated number of infected and the cumulated number of deaths. It is worthy of notice that the order of magnitude for the simulated infection cases substantially differs from that one of official data, while it is the same for the number of deaths. As already anticipated, the absence of an adequate tests sampling, especially in the first wave, has led to an unreliable number of infected recorded by the official institutions. Through the simulations it has been verified that, in order to obtain a comparable total number of deaths (about 39,000 simulated against the 35,000 real, from December 28th, 2019, until June 25th, 2020), it is necessary a number of circulating infected individuals running into the millions. This finding confirms the hypothesis, already discussed in Section 3, that the official data about infected was heavily underestimated. However, in terms of relative distribution of cases in the various regions, the comparison of the two chromatic maps in the first row of Figure 6 is quite good. The same holds also for the analogous comparison in Figure 7, where the simulated distribution of deaths among the Italian territories correctly identifies the northern regions as the most damaged, as in reality. The apparent discrepancy concerning the central and southern regions, which in the *status quo* simulation registered fewer deaths than in the real cases, can be explained recalling that the main approximation of the present model lies on the fact that each agent is representative of about 1,000 individuals. Therefore, the epidemic behaviour in regions with less than 1,000 deaths cannot be properly captured by the simulations, which return a null result.

The next step is to compare the *status quo* scenario with other alternative zone-based scenarios, i.e. M, H, V and RI. Looking again at Figure 6 and Figure 7, second and third rows, it can be noticed the expected increase in the number of infected and dead people due to the lower restrictions. As summarized in Table 4, the increase is about 20% on average for the total number of infected and 25% for the total number of deaths for the whole Italian territory. The increase for the most damaged region, i.e. Lombardy, goes from 20% to 31% for both infected and dead people. This suggests that, even if solutions with partial lockdowns are of course preferable from the socio-economic point of view, the increment in terms of loss of human life is not negligible with respect to the total lockdown, and should be carefully evaluated. On the other hand, there are no relevant differences in the results of the four indicator-based scenarios. Therefore, the RI scenario can be considered as the best solution, since is the one implying less restrictions, i.e. fewer regions in lockdown.

Pluchino et al. (2021) already demonstrated the effectiveness of RI as a good indicator of virus spreading and consequences on the population, by correlating its regional values with real data of the first wave of the pandemic. Here, we can try to support this result by simulating two new scenarios with RI = 1 for all regions, i.e. with the same *a-priori* risk for the whole territory, and showing that the results are not compatible with real data. Scenario N1 simulates virus spreading in absence of lockdown while scenario S1 simulates the same national lockdown of the *status quo*. As expected, a huge increase in the number of infected and deaths (more than 100%) with respect to the simulated *status quo* is observed, particularly in scenario N1. This suggests that the consequences in terms of infected and deaths would have been much worse if *a-priori* conditions would have no influence on virus spreading. Moreover, in these scenarios damages would be more uniformly distributed over the Italian territory than in reality, without substantial differences between northern and central-southern regions.

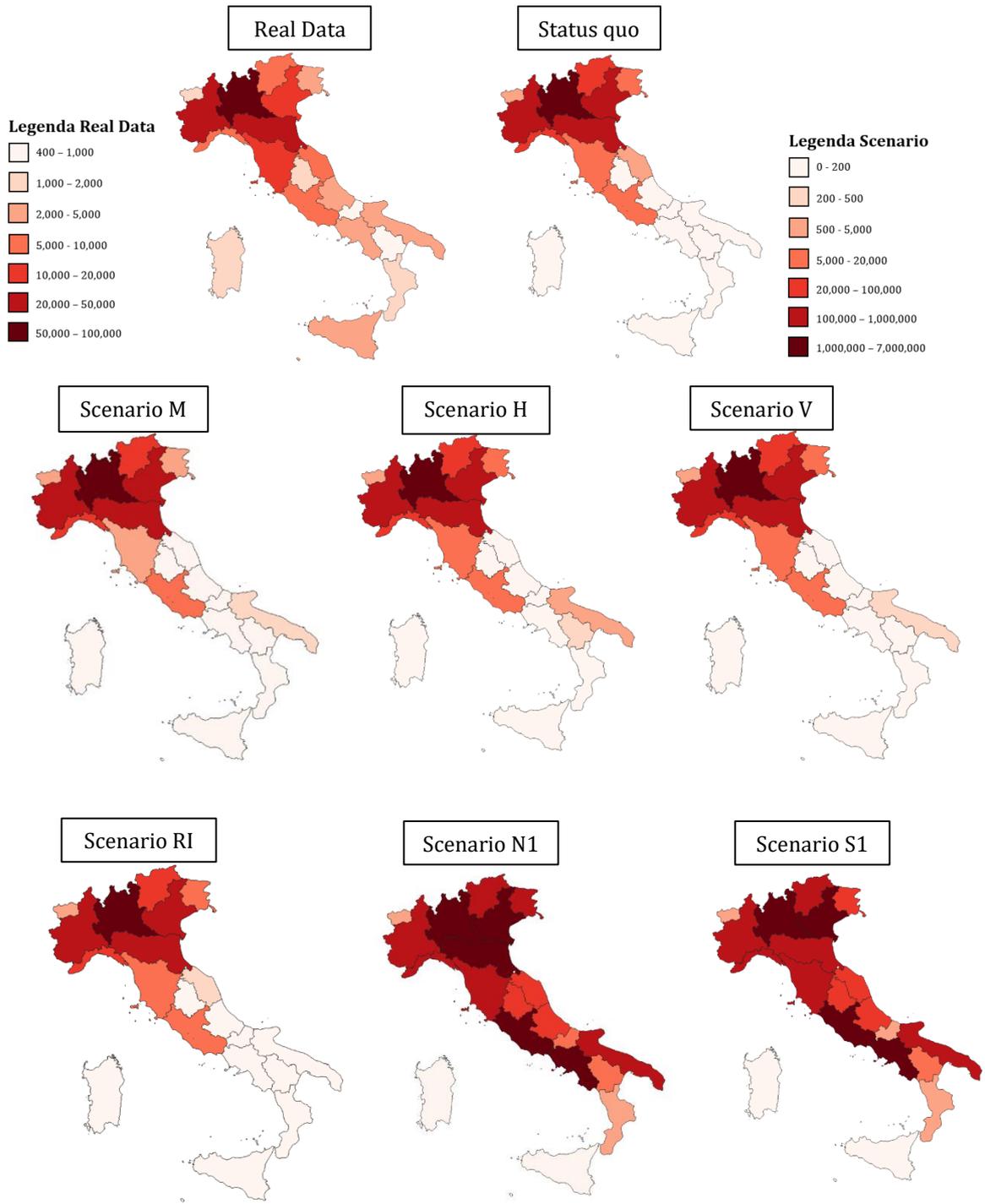

*Figure 6 - Distribution of number of infected for each scenario*

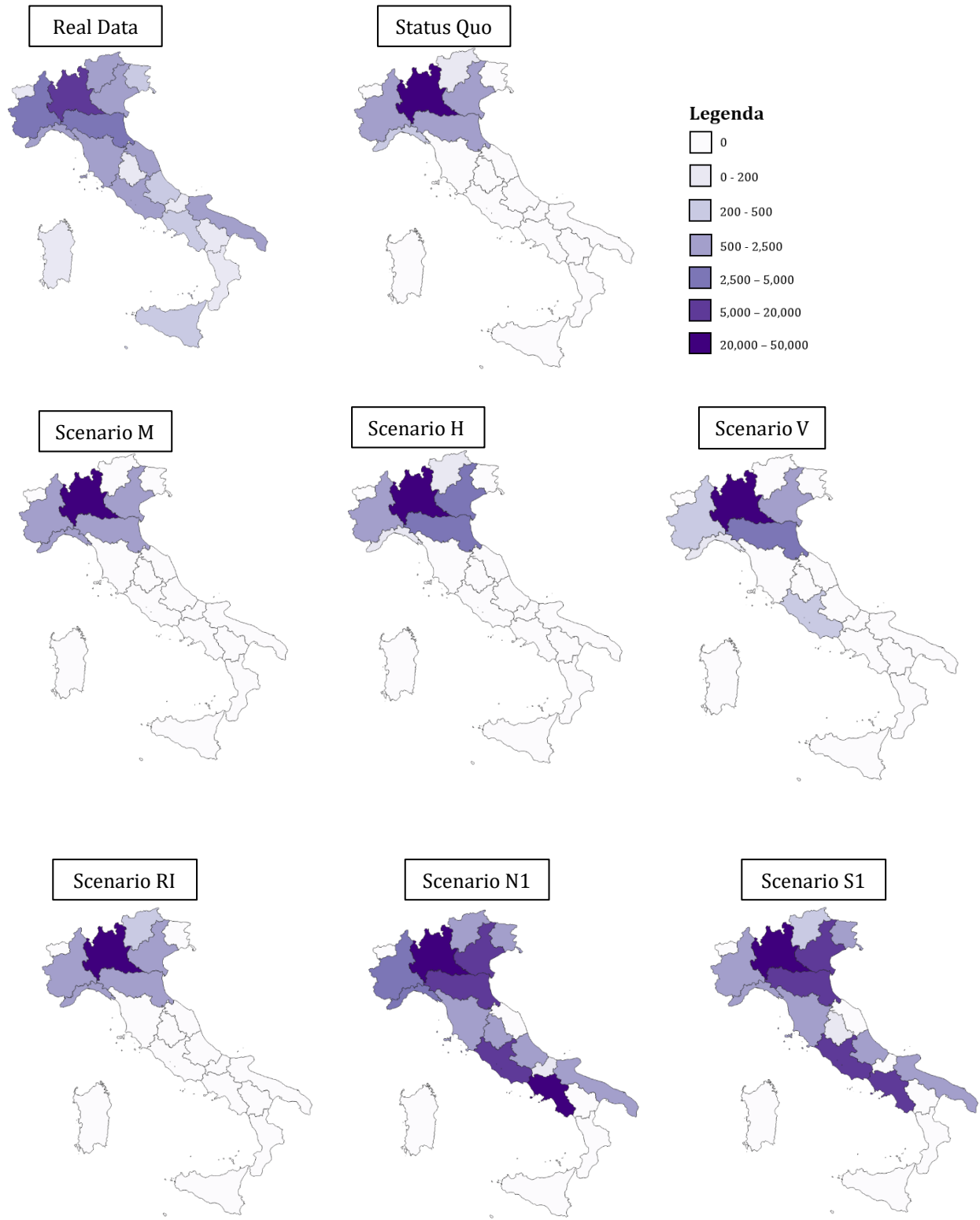

*Figure 7 - Distribution of number of deaths for each scenario*

| SCENARIO | Number of region for each zone | | | % increment of infected with respect to SQ | | % increment of deaths with respect to SQ | |
|---|---|---|---|---|---|---|---|
| | ZONE 1 | ZONE 2 | ZONE 3 | Total | Lombardy | Total | Lombardy |
| M | 6 | 9 | 5 | +22% | **+30%** | +28% | **+31%** |
| H | 4 | **10** | 6 | +17% | +25% | +30% | +30% |
| V | 12 | 7 | 1 | +22% | **+30%** | +18% | +20% |
| RI | 13 | 6 | 1 | +21% | +28% | +26% | +30% |
| N1 | **20** | 0 | 0 | **+123%** | +1% | **+187%** | +31% |
| S1 | 0 | 0 | **20** | +106% | -12% | +156% | +18% |

*Table 4 - Summary of the results obtained for each scenario*

In terms of policy implications, the main result is that differentiated mobility restrictions for the different regions are a suitable solution to limit virus spreading while reducing the overall impact on the economy. This is in line with the policies adopted by the national Government for the second wave that defined three zones (red, orange and yellow) based on multiple healthcare indicators, which depend both on the ability of each regions to cope with the virus spreading, and real-time data based on a continuous monitoring. However, our model suggests solutions that could be applicable for any sanitary emergency, regardless of real-time data, which could be, as in this case, naturally biased. In this respect, the ABM could be used to set the initial mobility restrictions, which should be updated according to the dynamic conditions linked to the virus spreading.

In other words, the model could be considered a decision-support tool for any strategic plan to contrast pandemics based on respiratory diseases, allowing a classification of regions based on *a-priori* data that could be regularly updated to have an up-to-date risk assessment for each region and know in advance the impact of different mobility restrictions strategies. This is particularly important and needed, given the unpreparedness of different countries to cope with the virus and, in the case of Italy, the lively debate around the outdated pandemic plan[9].

Moreover, since the model predicts the impact of the reduction of the radius of trips on the virus spreading, it could be also used to simulate targeted policies based on municipal, regional or national mobility. In the performed simulations, a lower radius for trips (i.e. short trips) reduces the risk of contagion because of a lower probability of getting in touch with other people, and this suggests that local policy-makers should guarantee adequate accessibility to essential services on short distances during pandemics and promote the use of sustainable transport modes to reach them, also in the view of the effects of pollution on the transmission of respiratory disease, proven also in the case of COVID-19 (Hensher, 2020, Gutiérrez et al. 2020). The Italian government is following this path, implementing national policies for the purchase of non-polluting vehicles (e.g. "Buono mobilità"), and promoting the construction of infrastructures dedicated to active mobility[10,11].

---

[9] https://www.theguardian.com/world/2020/aug/13/italy-pandemic-plan-was-old-and-inadequate-covid-report-finds

[10] https://www.gazzetta.it/bici/03-11-2020/milano-50-chilometri-nuove-piste-ciclabili-l-obiettivo-2020-390550946007.shtml

[11] https://www.gazzettaufficiale.it/atto/serie_generale/caricaDettaglioAtto/originario?atto.dataPubblicazioneGazzetta=2020-09-05&atto.codiceRedazionale=20A04737&elenco30giorni=false

# 5. Conclusion

In this paper an ABM model for dynamically simulating the impact of mobility restrictions during COVID-19 pandemics in Italy is presented. Different mobility limitation scenarios have been simulated with the aim of suggesting possible policy measure to limit the virus spreading. The scenario construction is based on the assignment of different mobility restrictions: no mobility restrictions at all, total mobility restrictions or different mobility restriction corresponding to 3 zones according to different parameters (i.e. M, H, V and RI). The main results show that assigning an *a priori* regional risk allows to adopt policies of localized restrictions that maintain almost the same effectiveness as a complete closure, allowing the opening of a greater number of economic activities and a greater mobility.

In this second wave, the Italian government decided a similar zonation, classifying Italian regions into three risk areas based on the progressive gravity of health emergency.

These government measures consider only sanitary parameters and are suitable for real-time management of the health emergency. However, at the beginning of the pandemic, the adoption of an adequate pandemic plan could have led to less drastic economic consequences (Haug et al., 2020).

In this respect, the model proposed in this paper aims at providing useful suggestions to contrast epidemic emergency in the context of a preliminary strategic plan. The reproducibility of the model and its scalability to different territorial contexts makes it a tool able to provide valuable information for government agencies to undertake the proper interventions in the event of a pandemic diffusion. In this respect, as future research, it should be tested in other contexts where the virus spreading followed different patterns. Finally, the ABM can also be adapted to other health emergencies caused by respiratory diseases.